\documentclass[prl,twocolumn,showpacs,superscriptaddress,nobalancelastpage]{revtex4}
\usepackage{graphicx}
\newcommand{\eq}[1]{(\ref{eq:#1})}

\begin{document}
\title{\bf Polarization qubit phase gate in driven atomic media}

\author{Carlo Ottaviani} 
\author{David Vitali} 
\affiliation{INFM and Dip. di Fisica Universit\`a di Camerino, I-62032 Camerino, Italy}
\author{Maurizio Artoni} 
\affiliation{INFM and Dip. di Chimica e Fisica dei Materiali, Via Valotti 9, 
I-25133 Brescia, Italy}
\affiliation{European Laboratory for Non-linear Spectroscopy, via
N. Carrara, I-50019 Sesto F.no, Italy}
\author{Francesco Cataliotti}
\affiliation{European Laboratory for Non-linear Spectroscopy, via
N. Carrara, I-50019 Sesto F.no, Italy}
\affiliation{INFM and Dip. di Fisica Universit\`a di Catania, via S. Sofia 64, 
I-95124 Catania, Italy}
\author{Paolo Tombesi}
\affiliation{INFM and Dip. di Fisica Universit\`a di Camerino, I-62032 Camerino, Italy}

\begin{abstract}
We present here an all--optical scheme for the experimental realization of a quantum phase gate. It is based on
the polarization degree of freedom of two travelling single photon wave-packets and exploits giant Kerr
nonlinearities that can be attained in coherently driven ultracold atomic media.
\end{abstract}

\pacs{03.67.-a, 42.65.-k, 42.50.Gy}

\maketitle

Photons are ideal carriers of quantum information as they travel at the speed of light and are negligibly
affected by decoherence. In fact, quantum key distribution \cite{qkd} and quantum teleportation
\cite{telep1,telep2} have been demonstrated using either single photon pulses, which encode the quantum
information in the photon polarization~\cite{qkd,telep1}, or squeezed light encoding the information in the field
quadrature \cite{telep2}. The use of photons has also been suggested for quantum computation schemes even 
though the absence
of significant photon-photon interactions becomes an obstacle toward the realization of efficient 
\textit{quantum gates}.
Two different ways have been proposed to circumvent this problem, namely, linear optics quantum
computation~\cite{klm} and nonlinear optical processes that involve few photons. While one is a probabilistic
scheme implicitly based on the nonlinearity hidden in single-photon detectors, the other is based on the
enhancement of photon-photon interaction achieved either in cavity QED configurations \cite{turch,arno,eke} or 
in dense atomic
media exhibiting electromagnetically induced transparency (EIT) \cite{nature}.

Single qubit gates and one universal two-qubit gates are required for implementing universal quantum computation.
The prototype optical implementation of a two-qubit gate is the quantum phase gate (QPG) in which one qubit gets
a phase conditional to the other qubit state according to the transformation $|i\rangle _{1}|j\rangle _{2}
\rightarrow \exp\left\{i \phi_{ij} \right\}|i\rangle _{1}|j\rangle _{2} $ where $\{i,j\}=0,1$ denote the logical
qubit bases. This gate becomes universal when $\phi=
\phi_{11}+\phi_{00}-\phi_{10}-\phi_{01}\neq 0 $ 
\cite{turch,lloyd}. 

Partial demonstrations of an optical QPG
have been already performed. A conditional phase shift $\phi \simeq 16^{\circ}$ between two frequency-distinct
cavity modes that experience an effective cross modulation mediated by a beam of Cs atoms has first been
measured nearly a decade ago~\cite{turch}. The complete truth table of a QPG has not been determined as yet and
an attempt in this direction has been made only very recently~\cite{resch} whereby a conditional phase shift $\phi
\simeq 8^{\circ}$ has been obtained between weak coherent pulses exploiting second-order nonlinearities in a
crystal. This experiment however does not seem to demonstrate a {\em bona fide} QPG as $\phi$ depends on the input
states and the gate works only for a restricted class of inputs. A
phase-tunable {\em mixed} QPG between a two-level Rydberg atom and the two lowest Fock states of a high-Q
microwave cavity has also been demonstrated~\cite{arno}.

A complete demonstration of a fully optical QPG is still lacking and we here envisage a new scheme
for the realization of such a logic gate. Our proposal relies on the polarization degree of freedom of two
travelling single-photon wave-packets and exploits the giant Kerr nonlinearities that can be observed in dense
atomic media under EIT \cite{ima}. A two-qubit gate for travelling photon qubits 
is useful not only for optical implementations of quantum computation, but also 
for quantum communication schemes. For example, perfect Bell-state discrimination 
for quantum dense coding and teleportation becomes possible if a QPG with a conditional phase 
shift $\phi = \pi$ could be used~\cite{prlno}. 

In our proposal the two qubits are a
\textit{probe} and a \textit{trigger} polarized single--photon wave--packet 
\begin{equation}
|\psi_{i}\rangle  = \alpha_{i}^{+}|\sigma^{+}\rangle _{i}+\alpha_{i}^{-}|\sigma^{-}\rangle _{i}, \quad i=
\{P,T \}
\end{equation}
which can be written in general as a superposition of two circularly polarized states
\begin{equation}
|\sigma^{\pm}\rangle _{i}=\int d\omega \ \xi_{i}(\omega) \hat{a}_{\pm}^{\dagger}(\omega) |0\rangle \label{base}
\end{equation}
where $
\xi_{i}(\omega) = \left(\tau_{i}^{2}/2\pi\right)^{1/4}
\exp\left\{-\tau_{i}^{2}\left(\omega-\omega_{i}\right)^{2}/4\right\}
$
is the  frequency distribution of the incident wavepackets centered on $\omega_{i}$ and with a time duration
$\tau_{i}$. In the interaction region of length $l$ 
the electric field operator undergoes the following transformation
\begin{equation}
\hat{a}_{\pm}(\omega) \rightarrow \hat{a}_{\pm}(\omega) \exp\left\{i \frac{\omega}{c} \int_{0}^{l}dz \
n_{\pm}(\omega,z)\right\}, \label{phase} \label{eq:a}
\end{equation}
where $n_{\pm}$ is real part of the refractive index which depends also on $z$ when cross--phase modulation is
present. Inserting \eq{a} into~(\ref{base}) and assuming that the refractive index varies slowly over the
bandwidth of the wavepackets, one gets
\begin{eqnarray}
|\sigma^{\pm} \rangle _{i} \rightarrow e^{-i \frac{\omega_{i}}{c}\int_{0}^{l}dz \ n_{\pm}(\omega_{i},z)}
|\sigma^{\pm} \rangle _{i}
 \equiv e^{-i\phi_{\pm}^{i}}|\sigma^{\pm} \rangle _{i}
\label{eq:squb}
\end{eqnarray}
yielding a two-qubit gate in the form,
\begin{equation}
|\sigma^{\pm} \rangle _{P} |\sigma^{\pm} \rangle _{T} \rightarrow e^{-i\left(\phi_{\pm}^{P}+\phi_{\pm}^{T}\right)}
|\sigma^{\pm} \rangle _{P} |\sigma^{\pm} \rangle _{T}.
\label{eq:tqub}
\end{equation}
This becomes a universal QPG~\cite{turch,lloyd} provided the conditional phase shift
\begin{equation}
 \phi =
\left(\phi_{+}^{P}+\phi_{-}^{T}\right)
- \left(\phi_{-}^{P}+\phi_{-}^{T}\right)
+ \{+ \longleftrightarrow - \}\neq 0.
\end{equation}
The two--qubit gate (\ref{eq:tqub}) 
could be implemented in a magnetically confined cold sample of $^{87}$Rb atoms where two weak and
well stabilized probe and trigger light beams exhibit a strong cross--Kerr effect in the M configuration that is
schematically described in Fig.~1. A $\sigma^{+}$ polarized probe couples the excited state $|2\rangle$ to the
ground $|1\rangle$ where all the atomic population is initially trapped. The other Zeeman split ground state
$|3\rangle $ is coupled to level $|4\rangle$ by a $\sigma^{-}$ polarized \textit{trigger} beam and to the excited
state $|2\rangle $ by an intense $\sigma^{-}$ polarized \textit{pump}. A fourth $\sigma^{-}$ polarized
\textit{tuner} beam couples level $|4\rangle $ and a third ground-state sublevel, $|5\rangle $. Owing to the
tuner, the trigger group velocity can be significantly slowed
down similarly to what happens to the probe. 
This represents an essential improvement over the N scheme of Ref.~\cite{ima} which does not involve the
tuner and where the trigger pulse, which is not slowed down, leads to a group velocity mismatch that significantly
limits the achievable nonlinear shifts \cite{harhau,luima}. We anticipate that in the present M scheme the group
velocity mismatch can instead be reduced to zero and the cross--Kerr nonlinearity made large enough to yield
cross--phase shift values of the order of $\pi$.
Phase--gating is realized when only one of the four possible probe and trigger polarization configurations in
\eq{tqub} exhibits a strong nonlinear cross--phase shift. 
For both $\sigma^{-}$ polarized probe and trigger it can be seen, in fact, that 
for not too large detunings there is no sufficiently close excited state to couple
level $|1\rangle$ to and no population in $|3\rangle$ to drive the relevant trigger transition. Likewise for a
$\sigma^{-}$ polarized probe and a $\sigma^{+}$ polarized trigger. In either case probe and trigger only acquire
the trivial vacuum phase shift $\phi_{0}^{i}=k_{i}l = \omega_{i} l /c$. When both probe and trigger are instead
$\sigma^{+}$ polarized, the former, subject to the EIT produced by the $|1\rangle$--$|2\rangle$--$|3\rangle$
levels $\Lambda$ configuration~\cite{eit,slow}, acquires a non trivial phase shift $\phi_{\Lambda}^{P}$ which can
be evaluated by neglecting trigger and tuner altogether, while the latter, off any close resonant level, acquires
again the vacuum shift $\phi_{0}^{T}$. Finally, for a $\sigma^{+}$ and $\sigma^{-}$ polarized probe and trigger, 
the two pulses will experience a substancial cross-Kerr effect acquiring nonlinear cross--phase 
shifts $\phi_{+}^{P}$ and $\phi_{-}^{T}$. We arrive then at the following QPG table
\begin{eqnarray}
&& |\sigma^{-} \rangle _{P} |\sigma^{-} \rangle _{T} \rightarrow e^{-i\left(\phi_{0}^{P}+\phi_{0}^{T}\right)}
|\sigma^{-} \rangle _{P} |\sigma^{-} \rangle _{T} \label{eq:gate1} \\
&&|\sigma^{-} \rangle _{P} |\sigma^{+} \rangle _{T} \rightarrow e^{-i\left(\phi_{0}^{P}+\phi_{0}^{T}\right)}
|\sigma^{-} \rangle _{P} |\sigma^{+} \rangle _{T} \\
&&|\sigma^{+} \rangle _{P} |\sigma^{+} \rangle _{T} \rightarrow e^{-i\left(\phi_{\Lambda}^{P}+\phi_{0}^{T}\right)}
|\sigma^{+} \rangle _{P} |\sigma^{+} \rangle _{T} \label{++ps}\\
&&|\sigma^{+} \rangle _{P} |\sigma^{-} \rangle _{T} \rightarrow e^{-i\left(\phi_{+}^{P}+\phi_{-}^{T}\right)}
|\sigma^{+} \rangle _{P} |\sigma^{-} \rangle _{T} \label{+-ps}
\label{eq:gate4}
\end{eqnarray}
with a conditional phase shift given by,
\begin{equation}
\phi= \phi_{+}^{P}+\phi_{-}^{T}-\phi_{\Lambda}^{P}-\phi_{0}^{T}.
\end{equation}

\begin{figure}[t]
\includegraphics[width=3.45in]{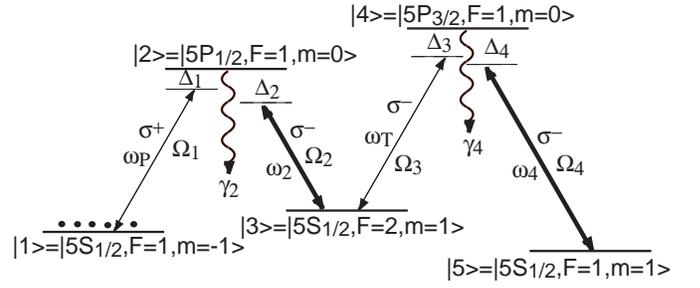}
\caption{ Polarization phase-gate in ultracold $^{87}$Rb. The
probe ($\omega_{P}$, $\Omega_{1}$) and trigger ($\omega_{T}$, $\Omega_{3}$) pulses impinging upon a Rb
sample in the presence of a strong pump ($\omega_{2}$, $\Omega_{2}$) and a tuner ($\omega_{4}$, $\Omega_{4}$)
realize the gating transformation (7--10). For a suitable choice of the four beam detunings ($\Delta_1, \Delta_2,
\Delta_3, \Delta_4$) and intensities, the $\sigma^+$ and $\sigma^-$ polarized probe and trigger can acquire a large
cross--Kerr phase modulation. The two excited states decay with rates $\gamma_2 \simeq \gamma_4 =\gamma =  2 \pi
\times 6$ MHz. } \label{fig1}
\end{figure}

Let us now explicitely evaluate the phase shift appearing in the required 
gate transformation (7--10). We start by
describing the system dynamics for the M configuration of Fig.~1 in terms of five coupled equations for the slowly
varying atomic amplitudes $c_{i}$~\cite{ima,haryama}, \textit{i.e.},
\begin{eqnarray}
i\dot{c}_{1} &=& -\frac{\Omega_{1}^{*}}{2}c_{2}  \label{ampli1} \\
i\dot{c}_{2} &=& \left(\Delta_{1}-i\frac{\gamma_{2}}{2}\right)c_{2}
-\frac{\Omega_{1}}{2}c_{1} -\frac{\Omega_{2}}{2}c_{3}\\
i\dot{c}_{3} &=& \Delta_{12}c_{3}
-\frac{\Omega_{2}^{*}}{2}c_{2} -\frac{\Omega_{3}^{*}}{2}c_{4}\\
i\dot{c}_{4} &=& \left(\Delta_{13}-i\frac{\gamma_{4}}{2}\right)c_{4}
-\frac{\Omega_{3}}{2}c_{3} -\frac{\Omega_{4}}{2}c_{5}\\
i\dot{c}_{5} &=& \Delta_{14}c_{5} -\frac{\Omega_{4}^{*}}{2}c_{4},  \label{ampli5}
\end{eqnarray}
where the relative detunings $\Delta_{12}=\Delta_{1}-\Delta_{2}$, $\Delta_{13}=\Delta_{12}+\Delta_{3}$ and
$\Delta_{14}=\Delta_{13}-\Delta_{4}$ are defined in terms of the detunings $\Delta_{1} = \omega_{21}-\omega_P$,
$\Delta_{2} = \omega_{23}-\omega_2$, $\Delta_{3} = \omega_{43}-\omega_T$, $\Delta_{4} = \omega_{45}-\omega_4$. We
here examine ultracold atomic samples at temperatures $T < 1$ $\mu$K so that Doppler broadenings and shifts can be
neglected. We assume that decay only occur from the two excited states $|2\rangle$ and $|4\rangle$
out of the system, with similar rates $\gamma_{2} \simeq \gamma_{4} 
= \gamma $ \cite{decay}. The pump and the tuner are taken as \textit{cw} light beams with constant Rabi frequencies
$\Omega_{2}$ and $\Omega_{4}$ while $\Omega_{1}$ and $\Omega_{3}$, referring to weak probe and trigger coherent
pulses, are space and time dependent Rabi frequencies. We determine the stationary state of
Eqs.~(\ref{ampli1})-(\ref{ampli5}) by assuming that most of the population remains in the initially populated
level $|1\rangle$; this occurs when the intensity of the pump is sufficiently larger than the probe intensity and
than the detunings as well, \textit{i.e.}, $ |\Omega_2|^2 \gg |\Delta_{12}(\Delta_1-i\gamma/2)|$. Under the
further assumption that the pump be stronger than the trigger as well, the stationary probe and trigger
susceptibilities can be rewritten as,
\begin{eqnarray}
&&\chi_{P}(z,t) \simeq \chi_{12}^{(1)}+\chi_{12}^{(3)}|E_{T}(z,t)|^{2} \label{eq:chipro}
 \\
&&\chi_{T}(z,t) \simeq \chi_{34}^{(3)}|E_{P}(z,t)|^{2}.
 \label{eq:chitri}
\end{eqnarray}
Here $E_{P}$ and $E_{T}$ are the probe and trigger electric fields while
\begin{eqnarray}
&&\chi_{12}^{(1)}=-\frac{N}{V}\frac{|\mu_{12}|^{2}}{\hbar \epsilon_{0}}
\frac{4\Delta_{12}}{|\Omega_{2}|^{2}} \label{eq:chi1pro}\\
&&\chi_{12}^{(3)}=\chi_{34}^{(3)}=\frac{N}{V}\frac{4|\mu_{12}|^{2}|\mu_{34}|^{2}}
{\hbar^{3} \epsilon_{0} |\Omega_{2}|^{2}}\left[\Delta_{13}-i\frac{\gamma}{2}-
\frac{|\Omega_{4}|^{2}}{4 \Delta_{14}}\right]^{-1}
\label{eq:chi3tri}
\end{eqnarray}
are respectively the linear and nonlinear susceptibilities given in terms of the dipole matrix elements $\mu_{12}$
and $\mu_{34}$ and atomic density $N/V$. These expressions yield previous results as limiting cases. The
third-order susceptibility for the N configuration assumed in~\cite{ima} is obtained when $\Omega_{4} =0$, while the trigger
susceptibility for the M configuration examined in~\cite{zub} obtains when $\Delta_{13}=0$.

The above results~\eq{chipro}-\eq{chi3tri} enable one to asses the group velocity mismatch between probe and
trigger. As pointed out in~\cite{luima}, the two group velocities have to be
comparable and small in order to achieve large cross-phase modulations. Unlike the six level
scheme studied in~\cite{kuri}, in which cross--phase modulation takes place in a
symmetric fashion so that the two group velocities are equal by construction, our present scheme is not
symmetrical and hence probe and trigger group velocities are not in general equal. The group velocities follow
from \eq{chi1pro} and \eq{chi3tri}
\begin{eqnarray}
&&v_{g}^{P} \simeq \frac{\hbar c \epsilon_{0}} {8\pi |\mu_{12}|^{2}\omega_{P} (N/V)} \times
\frac{|\Omega_{2}|^{2}}{1+\beta |\Omega_{3}|^{2}}\label{eq:vgpr}
 \\
&& v_{g}^{T} \simeq \frac{\hbar c \epsilon_{0}} {8\pi |\mu_{34}|^{2}\omega_{T} (N/V)} \times
\frac{|\Omega_{2}|^{2}}{\beta |\Omega_{1}|^{2}}, \label{eq:vgtr}
\end{eqnarray}
where
\begin{equation}
\beta = \frac{ \left(1+\frac{|\Omega_{4}|^{2}}{4\Delta_{14}^{2}}
\right)\left[\left(\Delta_{13}-\frac{|\Omega_{4}|^{2}}{4\Delta_{14}} \right)^{2}-
\frac{\gamma^{2}}{4}\right]}{\left [\left(\Delta_{13}-\frac{|\Omega_{4}|^{2}}{4\Delta_{14}}\right)^{2}+
\frac{\gamma^{2}}{4}\right]^{2}} .
\end{equation}
It follows that the two velocities can be made both small and equal by varying the probe and trigger relative
intensities and the parameter $\beta$. Because of the tuner, our present configuration enables one to further
control the group mismatch through $\beta$ which can be varied independently by adjusting the tuner intensity and
its relative detuning $\Delta_{14}$.

By comparing the qubits shifts in \eq{squb} with the solution
\begin{equation}
\varepsilon_{i}(z,t)=\varepsilon_{i}(0,t-\frac{z}{v_{g}^{i}}) \exp\left\{2\pi
ik_{i}\int_{0}^{z}dz'\chi_{i}(z',t)\right\}
 \label{eq:propa2}
\end{equation}
of the propagation equation~\cite{eit} for the slowly varying electric field amplitudes
$\varepsilon_{i}(z,t)$, where
$\chi_{i} \simeq (n_{i} -1)/2 \pi$ are given in \eq{chi1pro} and
\eq{chi3tri} and $v_{g}^{i}$ in \eq{vgpr} and
\eq{vgtr}, the phase in \eq{propa2} yields directly the required shifts for the phase--gating transformation
(7--10).
 The \textit{linear} phase--shift
$\phi_{\Lambda}^{P}$ acquired by a $\sigma^{+}$-polarized probe pulse moving in the $z$--direction across a sample
of optical thickness $l$ then becomes
\begin{equation}
 \phi_{\Lambda}^{P}= 
 k_{P} l \left \{ 1+2\pi \chi_{12}^{(1)} \right \}
\end{equation}
 while the \textit{nonlinear} shift is obtained when the last contribution on the right hand side of
\eq{chipro} is included. For a trigger Gaussian pulse~\cite{loud} of peak Rabi frequency 
$\Omega_{3}^{pk}$ and moving within the sample with
group velocity $v_{g}^{T}$, 
we arrive at an overall probe shift in the form
\begin{eqnarray}
&& \phi_{+}^{P}  =  \phi_{\Lambda}^{P} + 2\pi k_{P} \ \chi_{12}^{(3)}  
\int_{0}^{l}dz' \left|E_{T}(z^{\prime},t)\right|^{2} \nonumber \\
&=&   \phi_{\Lambda}^{P} +k_{P} l \frac{ \pi^{3/2} \hbar^{2}
|\Omega_{3}^{pk}|^{2}}{4|\mu_{34}|^{2}}
 \frac{{\rm erf}[\zeta_{P}]}{\zeta_P} {\rm Re} \chi_{12}^{(3)}
 \label{eq:xshift}
\end{eqnarray}
with $\zeta_{P}= \left(1-v_{g}^{P} / v_{g}^{T}\right) \sqrt{2} l / v_{g}^{P} \tau_{T}$ and where $\tau_{T}$ is
the trigger pulse time duration. By following the same procedure one has for the trigger phase--shift
\begin{eqnarray}
\phi_{-}^{T} = \phi_{0}^{T} +  k_{T} l \frac{ \pi^{3/2} \hbar^{2}
|\Omega_{1}^{pk}|^{2}}{4|\mu_{12}|^{2}}
\frac{{\rm erf}[\zeta_{T}]}{\zeta_T} {\rm Re} \chi_{34}^{(3)},
\end{eqnarray}
where $\zeta_{T}$ is obtained from $\zeta_{P}$ upon interchanging $P \leftrightarrow T$.

Large nonlinear shifts take place when probe and trigger velocities are very much alike, \textit{i.e.} when
$\zeta \rightarrow 0$ in which case the erf[$\zeta$]/$\zeta$ reaches the maximum value $2/\sqrt{\pi}$, and for
appreciably large values of the two nonlinear susceptibilities real parts. At the same time, their imaginary parts
have to be kept small so as to avoid absorption, which may hamper the efficiency of the gating mechanism. Assuming 
a perfect EIT regime for the probe, \textit{i.e.} $\Delta_{1}=\Delta_{2}=0$, 
it is easily seen from Eq.~(\ref{eq:chi3tri}) that 
one can attain imaginary parts that are two orders of magnitude smaller than their real parts for
suitable values of the tuner intensity and provided that trigger and tuner are both strongly detuned and by nearly
equal amounts, i.e. $\Delta_{3} \simeq \Delta_{4}$. Such a choice further leads to values of $\beta$ that yield
equal group velocities. By taking, e.g., 
$\Delta_{3} \simeq \Delta_{4}= 20  \gamma$ with 
$\Delta_{14} = 10^{-2} \gamma$, and $\Omega_{4} \simeq  \gamma$, $\Omega_{1} \simeq 0.08 \ \gamma$, 
$\Omega_{3} \simeq 0.04 \ \gamma$, 
$\Omega_{2}\simeq 2 \gamma$,  one has at typical densities of $N/V =3 \times 10^{13}$ cm$^{-3}$ group velocities
$v_{g}^{P} \simeq v_{g}^{T} \simeq 10$ m/s along with over 65 \% average transmission~\cite{tachyon} and a
conditional phase shift $\phi \simeq \pi$ over an interaction length  
$l \simeq 1.8$ mm. This set of Rabi frequencies corresponds to single photon 
probe and trigger pulses for tightly focused beams (several microns) 
with time duration $\sim 1$ $\mu$s. The non negligible absorption accompanying the nonlinear phase 
shift does not hinder the proposed QPG mechanism.
A demonstration of the proposed QPG may be done by using post-selection 
of single--photon coherent pulses instead of single photon wave-packets. 
In this case, the phase gating
mechanism described by Eqs.~(\ref{eq:gate1})-(\ref{eq:gate4}) is carried out by considering the
four possible configurations for the input polarizations, measuring the phase shifts with a Mach-Zender
interferometer set-up~\cite{resch}, and post-selecting only the events with a coincident detection of one photon
out of each probe and trigger pulse. Non negligible absorption implies then only a smaller value of probe and
trigger transmitted amplitudes with a concomitant lower probability (by $40\%$) 
to detect a two-photon coincidence between probe and trigger.

Laser pump intensity and frequency fluctuations may increase absorption and phase-shift fluctuations. 
The gate fidelity may then be hampered though in the proposed post-selection scheme, 
the fidelity is essentially affected only by the fluctuations of the shifts 
$\phi_{\Lambda}^{P}$, $\phi_{-}^{T}$ and $\phi_{+}^{P}$. On general ground one 
estimates that a $1\% $ intensity fluctuation yields an error probability of 
about $3 \% $ though relative detuning fluctuations of the order of $10^{-5}\gamma $ 
can make the error probability to become as large as $10 \% $ \cite{fluct}.
It is worthwhile to note that a classical phase gate could be implemented
by using more intense probe and trigger pulses. In fact, a conditional phase shift $\phi \simeq \pi$
could be achieved with the same atomic density but over a
shorter interaction length,  $l \simeq 10 \mu$m,
along with 80 \% average transmission, by choosing
$\Omega_{1} \simeq 1.4 \ \gamma$, $\Omega_{3} \simeq 0.16 \ \gamma$, 
$\Omega_{4} \simeq  \gamma$,
$\Omega_{2}\simeq 7 \ \gamma$ and by slightly decreasing the detunings $\Delta_3$ and $\Delta_4$.

We here propose in conclusion a feasible scheme for an all-optical quantum phase gate that uses
travelling single-photon pulses in which quantum information is encoded in the polarization degree of freedom.
Unlike a similar scheme already investigated in \cite{zub,green} and where the issue of the two
probe and trigger pulses group velocities mismatch was not addressed, we here observe that a $\pi$ phase
shift is obtained only when the probe and trigger group velocities are both small and almost equal. We show, 
within the framework of the present model, that this can be realized simply by tuning the 
frequencies and intensities of the four input light
beams. This way of achieving a zero group velocity mismatch has clear advantages over other schemes that have been
recently discussed \cite{luima,kuri}. The proposed scheme could be directly applied in fact to a magnetically
confined cold sample of $^{87}$Rb atoms and does not require a cold trapped mixture of two atomic species as
in~\cite{luima}, where the two species realizing a N and a $\Lambda$ scheme respectively, require an accurate
control of the atomic densities in order to get equal group velocities. The scheme studied instead in~\cite{kuri}
is symmetric for probe and trigger and therefore yields equal group velocities automatically. Yet, the initial
atomic population is here to be put in a Zeeman-split $m=0$ ground state sublevel which cannot be easily done in a
magnetically confined atomic sample requiring more sophisticated optical trapping techniques.

We acknowledge enlightening discussions with S. Harris, M. Inguscio and T. Arecchi regarding the pulse propagation
in ultracold rubidium samples. This work has been supported by the EU (Contract HPRICT1999-00111), the Italian
Ministry of University and Research (MURST-Integrated Actions) and the MIUR (PRIN 2001 \textit{Quantum
communications using slow light}).


\begin{thebibliography}{99}


\bibitem{qkd}N. Gisin {\it et al}.,
Rev. Mod. Phys. {\bf 74}, 145 (2002).
\bibitem{telep1}D.~Bouwmeester {\it et al}., Nature (London) {\bf 390},
575 (1997); D.~Boschi {\it et al}., Phys.  Rev.  Lett.  {\bf 80}, 1121 (1998).
\bibitem{telep2}A. Furusawa {\it et al}., Science, {\bf 282} 706
(1998).
\bibitem{klm}E. Knill {\it et al}., Nature
(London) {\bf 409}, 46 (2001).
\bibitem{turch}Q.A. Turchette {\it et al}.,
 Phys. Rev. Lett. {\bf 75}, 4710 (1995).
\bibitem{arno}A. Rauschenbeutel {\it et al}., Phys. Rev. Lett. {\bf 83},
5166 (1999).
\bibitem{eke}V. Giovannetti, {\it et al.}, Phys. Rev. A 62, 032306 (2000).
\bibitem{nature}M. D. Lukin and A. Imamo\u{g}lu, Nature (London)
{\bf 413}, 273 (2001) and references therein.
\bibitem{lloyd}S.~Lloyd, Phys. Rev. Lett. {\bf 75}, 346 (1995).
\bibitem{resch} See, e.g., K. J. Resch {\it et al}., Phys. Rev. Lett. {\bf 89},
037904 (2002).
\bibitem{ima}H. Schmidt and A. Imamo\u{g}lu, Opt. Lett. {\bf 21},
1936 (1996).
\bibitem{prlno}D. Vitali {\it et al}., Phys. Rev. Lett. {\bf 85}, 445
(2000).
\bibitem{harhau}S. E. Harris and L. V. Hau, Phys. Rev. Lett. {\bf 82},
4611 (1999).
\bibitem{luima}M. D. Lukin and A. Imamo\u{g}lu, Phys. Rev. Lett. {\bf 84},
1419 (2000).
\bibitem{eit}E. Arimondo, in {\it Progress in Optics}
XXXV, ed. by E. Wolf, (Elsevier, Amsterdam, 1996);
S. E. Harris, Phys. Today {\bf 50}, 36 (1997); M. O. Scully
and M. S. Zubairy, {\it Quantum Optics} (Cambridge University Press,
Cambridge, UK, 1997).
\bibitem{slow}L. V. Hau {\it et al}., Nature (London) {\bf 397}, 594
(1999); M. M. Kash {\it et al}., Phys. Rev. Lett. {\bf 82}, 5229 (1999).
\bibitem{haryama} S. E. Harris and Y. Yamamoto, Phys. Rev. Lett {\bf
81}, 3611 (1998).
\bibitem{decay}The assumption of ``external'' decays allows us to use 
\cite{ebe} the amplitude equations (12-16), upon which our treatment is based, instead of optical
Bloch equations. The use of Bloch variables along with ``internal'' decays
would make the treatment more involved and the basic physics underlying the
gating mechanism less transparent while reaching results unchanged in essence.
\bibitem{ebe}L. Allen and J. Eberly, {\it Optical Resonance and Two-level atoms} (Dover,
New York, 1987)
\bibitem{zub}M. S. Zubairy {\it et al}., Phys. Rev.
A {\bf 65}, 043804 (2002); A. B. Matsko {\it et al}., Opt. Lett.
{\bf 28}, 96 (2003).
\bibitem{kuri}D. Petrosyan and G. Kurizki, Phys. Rev. A {\bf 65},
033833 (2002).
\bibitem{loud}K. J. Blow {\it et al.}, Phys. Rev. A {\bf 42}, 4102 (1990).
\bibitem{green}A. D. Greentree {\it et al}., e-print archive
quant-ph/0209067.
\bibitem{tachyon} M. Artoni, {\it et al}.,
Phys. Rev. A {\bf 63}, 023805 (2001).
\bibitem{fluct}Yet, error probability as small as $1 \% $ or less are achieved when, e.g., 
$\Delta_{12}$ is stabilized at $10^{-6}\gamma $ or more. Such a requirement, though stringent, 
is within experimental reach provided all lasers are tightly 
phase-locked to each other.


\end{thebibliography}
\end{document}